\documentclass[11pt,a4paper]{elsarticle}
\usepackage{color}
\usepackage{lineno}
\usepackage{hyperref}
\usepackage{amsmath}
\usepackage{amssymb}
\usepackage{slashed}
\usepackage{graphicx}
\modulolinenumbers[5]

\makeatletter \def\ps@pprintTitle{%
 \let\@oddhead\@empty
 \let\@evenhead\@empty
 \def\@oddfoot{\centerline{\thepage}}%
 \let\@evenfoot\@oddfoot}
 \makeatother

\biboptions{sort&compress}
\journal{ }

\begin{document}

\begin{frontmatter}

\title{Non-Abelian T-Dualizing the Resolved Conifold with Regular and Fractional D3-Branes}
\author[main]{K. S. Kooner}
\ead{K.S.Kooner.745700@swansea.ac.uk}
\author[bmain,bbmain]{S. Zacar\'ias}
\ead{szacarias@fisica.ugto.mx}
\address[main]{Department of Physics, Swansea University, Singleton Park, Swansea SA2 8PP, United Kingdom}
\address[bmain]{Departamento de F\'{\i}sica, Divisi\'{o}n de Ciencias e Ingenier\'{\i}as, Campus Le\'{o}n, Universidad de Guanajuato, Loma del Bosque No. 103 Col. Lomas del Campestre, C.P. 37150, Le\'{o}n, Guanajuato, M\'{e}xico}
\address[bbmain]{Department of Nuclear and Particle Physics, Faculty of Physics, University of Athens,  Athens 15784, Greece}

\begin{abstract}
In this paper we obtain new solutions of Type IIA and massive Type IIA supergravity.
These solutions are the result of implementing a non-abelian T-duality along the internal $SU(2)$ isometries of several D3-brane configurations on the resolved conifold, studied by Pando Zayas and Tseytlin. We first study the pure NS resolved conifold solution, then we add fluxes by placing a stack of D3-branes at the tip of the resolved conifold and finally we consider the system of regular and fractional D3-branes at the tip. We present the non-abelian T-duals associated with these backgrounds and study their geometries and fluxes. We briefly comment on some field theory features by studying couplings and the central charge of the dual field theory. We also analyze the supersymmetry of the dual solutions and show that for the system of only D3 branes the duality defines a map between backgrounds with $SU(3)$ and orthogonal $SU(2)$ structures.
\end{abstract}

\end{frontmatter}

\section{Introduction}

The AdS/CFT correspondence \cite{hep-th/9711200, hep-th/9802109, hep-th/9802150} has provided a powerful tool in the understanding of strong coupling dynamics in gauge theories.
The original and well-known example of the correspondence was the relation between Type IIB string theory on $\textrm{AdS}_5 \times S^5$ and a 4d $\mathcal{N}=4$ conformal SYM gauge theory, and subsequent work has attempted to find correspondences with more relevant gauge theories, in particular those that are non-conformal and less supersymmetric.
Along these lines, it has been of interest to study backgrounds that have more general internal-space structures, of which the conifold has been of central importance, being a straightforward way to break the maximal symmetry of $S^5$. The conifold is a Calabi-Yau 3-fold whose base manifold is $S^2 \times S^3$, and which has a singularity at its conical tip.
 The first system studied in this background was presented by Klebanov and Witten (KW) in \cite{hep-th/9807080},  in which a configuration of $N$ D3-branes placed at the tip of the conical singularity was considered.
 It was found that Type IIB string theory on $AdS_5\times T^{1,1}$ is dual to an $\mathcal{N}=1$ $SU(N) \times SU(N)$ conformal gauge theory with bifundamental matter fields.
Whilst still preserving $\mathcal{N}=1$ SUSY, conformal symmetry may be broken by adding $M$ fractional D3-branes to the KW model, modifying the gauge group to $SU(N+M) \times SU(N)$, as demonstrated by Klebanov and Tseytlin (KT) in \cite{Klebanov:2000nc}.
This model was seen to undergo a succession of Seiberg duality operations.
However, the presence of a singularity in the IR meant that the observed duality cascade could only be trusted in the UV.
This problem was overcome by Klebanov and Strassler (KS) by replacing the singular conifold by its smooth deformation \cite{hep-th/0412187}.
In fact this solution was found to lie on a class of solutions called the baryonic branch of Klebanov-Strassler (KS+bb) \cite{hep-th/0412187}, which interpolates between the KS solution, where regular and fractional D3-branes are placed at the tip of the deformed conifold, and the CVMN solution \cite{ hep-th/9707176, hep-th/0008001}, which is the near-brane system of D5-branes wrapping the $S^2$ of $T^{1,1}$.
This class of backgrounds is conjectured to be dual to non-conformal $\mathcal{N}=1$ confining gauge theories. For instance,  the KS solution undergoes a cascade of Seiberg dualities to a pure $SU(M)$ gauge theory.
The singularity at the tip of the conifold may however be smoothed out in another way: by replacing the conifold with its smooth resolution.
In this paper we are interested in studying D-brane configurations in this smooth background and constructing new supergravity solutions by means of string dualities.

Dualities have proved an important tool in the understanding of several aspects of string theory, and one of the best understood and most insightful is T-duality.
This symmetry appears due to the ability of closed strings to wind around circles, and has no analogue in ordinary quantum field theories.
From the space-time perspective, T-dualities modify the target geometry in such a way that the winding modes of the string coincide with the momentum modes in the dual picture and vice versa.
In this vein, it is natural to ask if this symmetry can be extended to more general compact spaces, and in particular we are interested in generalizing this idea from strings winding around circles to spheres.
This will involve generalizing the T-duality Buscher procedure \cite{Buscher:1987qj} from operating on an abelian $U(1)$ subgroup to a non-abelian $SU(2)$ subgroup \cite{hep-th/9210021}.
This work was initiated in \cite{Sfetsos:2010uq}, where the dualization of the maximally supersymmetric $\textrm{AdS}_5 \times \textrm{S}^5$ solution along the $SU(2)$ isometry of the internal space was performed.
The resulting dual model was found to be a solution of Type IIA supergravity, whose M-theory lift was related to gravity duals of $\mathcal{N} = 2$ superconformal theories \cite{Gaiotto:2009gz}.
In light of this result the non-abelian T-duality (NATD) procedure was applied to the KW, KT, KS, CVMN and KS+bb solutions in order to generate new supergravity duals and their associated field theories at  strong coupling \cite{Itsios:2013wd}.
These new backgrounds were found to exhibit interesting properties such as confinement, Seiberg-like dualities, and the existence of domain walls; it was also found that certain field theory quantities like the entanglement entropy and central charge are left invariant under NATD up to a constant volume factor.
Other backgrounds that have been recently studied include the $Y^{p,q}$ class of geometries, where the NATD background was found to be a class of Type IIA solutions preserving $\mathcal{N}=1$ supersymmetry, explicitly found to support an $SU(2)$ structure \cite{arXiv:1408.6545}; the ABJM model of $\textrm{AdS}_4$ backgrounds, whose dual is an $\mathcal{N}=2$ SUSY background whose CFT is a doubling of the original ABJM gauge groups \cite{Lozano:2014ata}; and particularly interestingly the previously only known SUSY $\textrm{AdS}_6$ model of \cite{hep-th/9905148} in which the NATD procedure was used to generate a new supersymmetry preserving $\textrm{AdS}_6$ solution of Type IIA supergravity \cite{Lozano:2012au}, this being the uplift of the supersymmetric vacuum of $F(4)$ gauged supergravity \cite{Jeong:2013jfc}. Remarkably, it was also observed in \cite{Jeong:2013jfc} that Type IIB and Romans' theories are linked by the gauging of SU(2) isometries. 

A large class of new $\textrm{AdS}_5$ Type IIB solutions was presented in \cite{Macpherson:2014eza}, whose dual CFTs were also studied, and a sample of the large body of work constructing new supergravity solutions using non-abelian T-duality, as well analyses of the dual CFTs, can be found in the papers \cite{Gaillard:2013vsa, Macpherson:2013zba, Lozano:2013oma, Zacarias:2014wta, Pradhan:2014zqa, Caceres:2014uoa, Araujo:2015npa, Bea:2015fja}.

In \cite{hep-th/0010088} it was shown that a warped geometry can be obtained by considering a system of regular and fractional D3-branes placed at the tip of the resolved conifold.
The dual field theory of this system was conjectured to lie on the mesonic branch \cite{Aharony:2000pp} and shown to be non-supersymmetric \cite{Cvetic:2000db}.
In the case of only regular D3-branes the field theory dual conspires to give a VEV to the bifundamental fields such that the operator $\mathcal{U}\sim \textrm{Tr}(B_{i}^{\dag}B_{i}-A_{i}^{\dag}A_{i})$ 
is non-zero \cite{Klebanov:1999tb}. 
Here, we construct the T-duals\footnote{Henceforth, T-duality will specifically mean non-abelian T-duality and the standard T-duality will always be referred to as abelian T-duality.}
of these solutions by following the procedure outlined in \cite{Itsios:2013wd}. However, the results for a general Type II background were presented in \cite{Kelekci:2014ima} and one may simply compare input backgrounds and write the T-dual output.
For the system of regular and fractional D3-branes we shall see that the resulting dual solution is a solution of massive Type IIA supergravity with a well quantized Romans mass.
In the case of only regular D3-branes we shall see that the dual background defines an $SU(2)$ orthogonal structure.
We shall also see how the T-dual solutions obtained here behave in the IR and UV limits and how they connect to the T-duals of the other conifold-like solutions presented in \cite{Itsios:2013wd}.
And finally, we will study the Page charges and central charge before and after the NATD with the aim of examining how they behave under the duality.

The organization of the paper is as follows: in section \ref{secNATD} we begin by giving a brief overview of the NATD procedure.
In section \ref{resolvedempty}, before we begin to add fluxes to the resolved conifold, we apply the NATD procedure to the pure NS solution.
In section \ref{sec:regularD3} we add regular D3-branes smeared over the tip of the resolved conifold, and in section \ref{sec:reg+fracD3} we additionally add fractional D3-branes at the tip of the resolved conifold and calculate the corresponding NATD solutions, all of which are found to be solutions of Type IIA supergravity, and in the case of the PT solution proper of massive Type IIA supergravity.
We shall verify, using the language of G-structures, the non-supersymmetric nature of this solution, which straightforwardly will serve to verify that the system of regular D3-branes is an $\mathcal{N}=1$ vacuum, defining an SU(3) structure which will become an orthogonal $SU(2)$ structure after NATD.
In section \ref{sec:probes} we analyze the central charge and Page charges of this solution, and we close in section \ref{cremarks} with some concluding remarks.

\section{Non-abelian T-duality: the general strategy}\label{secNATD}
In this section we shall give an overview of the non-abelian T-duality technique; we refer 
the interested reader to \cite{Itsios:2013wd} for more in-depth details, complemeted in part with \cite{Macpherson:2014eza}.

Let us start by considering an $SU(2)$ invariant background in such a way that the metric can be written in the form
\begin{gather}
ds^2=G_{\mu\nu}(x)dx^{\mu}dx^{\nu}+2G_{\mu i}(x)dx^{\mu}L^{i}+G_{ij}(x)L^{i}L^{j},
\label{metricdec}
\end{gather}
where $L^{i}=-i\textrm{Tr}(g^{-1}dg)$ are the left-invariant Maurer-Cartan 1-forms obeying the relationship $dL^i = \frac{1}{2} f^i_{jk} L^j \wedge L^k$.
Here the directions $i,j=1,2,3$ are $SU(2)$-dependent, which can be parametrized in terms of Euler angles $\theta,\phi, \psi$, whereas 
the coordinates over the indices $\mu,\nu=1,2,...,7$ are spectator coordinates under the duality transformation.
In the general case the seed background also contains a dilaton field $\Phi$ and a non-zero 2-form $B$, which we similarly deconstruct as:
\begin{gather}\label{Bdeconstruction}
	B = \frac{1}{2} B_{\mu \nu}(x) dx^\mu \wedge dx^\nu + B_{\mu i} dx^\mu \wedge L^i + \frac{1}{2} B_{ij}(x) L^i \wedge L^j.
\end{gather}
The action for the non-linear sigma model is
\begin{gather}
S=\int d^{2}\sigma\left(Q_{\mu\nu}\partial_{+}x^{\mu}\partial_{-}x^{\nu}+Q_{\mu i}\partial_{+}x^{\mu}L_{-}^{i}+Q_{i\mu}L_{+}^{i}\partial_{-}x^{\mu} +Q_{ij}L_+^{i}L_{-}^{j}\right),
\label{sigmam}
\end{gather}
where we have defined $Q=G+B$.
As in the abelian case, one may apply the gauging procedure by first introducing $SU(2)$-valued gauge fields in the usual way by replacing derivatives by covariant derivatives.
We then constrain the gauge fields to be pure gauge by adding to the action the term $-i\textrm{Tr}(vF)$, where $F = \partial_+ A_- - \partial_- A_+ - [A_+,A_-]$.
The T-dual model is then obtained by integrating out the gauge fields, leaving a Lagrangian dependent on $\theta,\phi,\psi$ and now the Lagrange multipliers $v_i$. We then gauge away this redundancy by setting three of the variables equal to zero, with different such gauge choices being locally diffeomorphic to one another. The simplest choice is $g = \mathbb{I}$.
We then get the dual action

\begin{gather}
\hat{S}=\int d^{2}\sigma\left(Q_{\mu\nu}\partial_{+}x^{\mu}\partial_{-}x^{\nu}+(\partial_{+}v_{i}+\partial_{+}x^{\mu}Q_{\mu i})(Q_{ij}+f_{ij}^{k}v_{k})^{-1}(\partial_{-}v_{j}-Q_{j\mu}\partial_{-}x^{\mu})\right),
\label{tduals}
\end{gather}
Then, from the above expression, one can easily read off the transformed NS fields $(G,B)$ in the dual background.
The dilaton field is also influenced by the duality transformation by a Jacobian factor in the measure due to the above manipulations. Explicitly it is
\begin{gather}
\hat{\Phi}=\Phi-\frac{1}{2}\det M.
\end{gather}
The NATD rules for the RR fluxes are obtained in a different manner. As in the abelian case, we find two sets of frame fields after dualization  which are related by a local 
Lorentz transformation $\Lambda$. The spinor representation, $\Omega$, of this Lorentz transformation tells us how these fluxes transform.  We find {\cite{Sfetsos:2010uq}
\begin{gather}
e^{\Phi_{\textrm{IIA}}}{\slashed F}_{\textrm{IIA}}=e^{\Phi_{\textrm{IIB}}}{\slashed F}_{\textrm{IIB}}\cdot \Omega^{-1},
\label{omegat}
\end{gather}
where $F_{IIA/IIB}$ are RR polyforms, which are bispinors under the Clifford map: ${\slashed F_p}=\frac{1}{p!}\Gamma_{\alpha_1...\alpha_p}F^{\alpha_1...\alpha_p}$.
It is worth noting that alternative derivations of eq. (\ref{omegat}) have been reported in the literature (see \cite{Kelekci:2014ima, Jeong:2013jfc, Gevorgyan:2013xka} for a sample of this work).

\section{The resolved conifold solution}\label{resolvedempty}\label{sec1}
In this section we shall obtain the dual background associated with the resolved conifold geometry by applying the NATD technique outlined in section \ref{secNATD}.
Let us first start by reviewing the resolution of the conifold singularity.

The resolution of the conifold singularity can be derived from the homogeneous equation that defines the singular conifold embedded in $\mathbf{C}^4$,
\begin{gather}
	\sum_{i=1}^{4}\omega^{2}_{i}=0,
	\label{conifold}
\end{gather}
and then making a change of variables to bring eq. (\ref{conifold}) into the form $xy-uv=0$. We then replace this equation by requiring non-trivial solutions of the system 
\begin{align}
	\begin{pmatrix}
	    z & u \\
	    v & y
	\end{pmatrix}
	\begin{pmatrix}
	    \xi_1 \\
	    \xi_2
	\end{pmatrix}
	=0 ,
\end{align}
where $\xi_i\neq 0$. At the tip of the conifold we find solutions given by $(\xi_1, \xi_2)$. Then by removing the invariance under $\xi_i \sim \lambda \xi_i$, with $\lambda \in \mathbb{C}$, one easily sees that this pair  actually describe an $S^2$.
The geometry of the resolved conifold may be obtained by imposing preservation of the Calabi-Yau structure. This was first presented in \cite{Candelas:1989js} and it is topologically an $\mathbf{R}^4$ bundle over $S^2$,
which is explicitly given by 
	\begin{equation}\label{dsresolved}
\begin{split}
		ds_{6}^2 = \kappa^{-1}(r) \, dr^2 &+ \frac{1}{6} r^2 \left( d\theta_1^2 + \sin^2 \theta_1 \, d\phi_1^2 \right) + \frac{1}{6}(r^2 + 6 a^2)\left( d\theta_2^2 + \sin^2 \theta_2 \, d\phi_2^2 \right) \\
																												  &+ \frac{1}{9} \kappa(r) r^2 \left( d\psi + \sum_{i=1}^2 \cos \theta_i \, d\phi_i \right)^2 ,
\end{split}
\end{equation}
where
\begin{equation}
	\kappa(r) = \frac{r^2 + 9 a^2}{r^2 + 6 a^2} .
\end{equation}
The resolved conifold metric is therefore a smooth $SU(2)_1\times SU(2)_2\times U(1)_{\psi}$ invariant manifold
defining a cone over the base space $S^2 \times S^3$.
	Note that at the tip the size of the $S^3$ parametrized by $(\theta_1,\phi_1,\psi)$ shrinks to zero whereas the $S^2$ $(\theta_2,\phi_2)$ has radius $a$, which is called the 
	resolution parameter. Note also that when we reduce the value of this parameter down to zero we recover the singular conifold geometry. In preparation for the NATD, we can choose frame fields for the metric in eq. (\ref{dsresolved}) as

\begin{gather}
	 e^{\theta_1 , \phi_1} = \lambda_1(r) \sigma_{\hat 1, \hat 2} , \quad
	 e^{1,2} = \lambda_2(r) \sigma_{1,2} , \quad
	 e^3 = \lambda(r) (\sigma_3 + \cos \theta_1 \, d\phi_1),
	\label{frames1}
\end{gather}
where
\begin{gather}
	\lambda_1^2(r) = r^2/6,\quad
	\lambda_2^2(r) = \left(r^2 + 6 a^2\right)/6 , \quad
	\lambda^2(r) = \kappa(r) r^2/9 ,
	\label{lambdas}
\end{gather}
and
\begin{gather}
	\sigma_{\hat 1} = d\theta_1 , \quad \sigma_{\hat 2} = \sin \theta_1 \,d\phi_1 ,
	\label{sigma_hat_defns}
\end{gather}
where the left-invariant 1-forms $\sigma_i$ are
\begin{gather}
\sigma_1 = \cos \psi \sin \theta_2 \, d\phi_2 - \sin \psi d\theta_2, \quad
		\sigma_2 = \sin \psi \sin \theta_2 \, d\phi_2 + \cos \psi d\theta_2, \nonumber\\
		\sigma_3 = d\psi + \cos \theta_2 \, d\phi_2 .
		\label{eulerforms}
\end{gather}
Note that the $\lambda$'s in eq. (\ref{lambdas}) have an explicit $r$-dependence that is non-uniform, whereas far away from the tip these running  $\lambda$'s all scale equally as $r^2$, which it should be noted is typically extracted from the $\lambda$'s in other papers.
Note also the asymmetry induced between $\lambda_1$ and $\lambda_2$ by the resolution parameter, which we will see again once we add fluxes to the geometry.


We now dualize this pure-NS solution along the $SU(2)_2$ isometry defined by the $\sigma_{i}' s$ in eq. (\ref{eulerforms}), with Lagrange multipliers $v_i$. We gauge fix by retaining the $\psi$ coordinate and killing one of the Lagrange multipliers, choosing $v_2 = 0$, and also defining $v_1 \equiv 2x_1$ and $v_3 \equiv 2x_2$. Following the conventions of \cite{Itsios:2013wd}, we deconstruct frame fields as
\begin{gather}
	e^A = e^A_\mu dy^\mu, \quad e^a = \kappa^a_i L^i + \lambda^a_\mu dy^\mu ,
	\label{NATD:vielbein_defn}
\end{gather}
where the indices run over $A=1,\dots,7$; $a,i = 1,2,3$ and $\mu=1,\dots,10$. We can then define the central quantities in the NATD procedure,
\begin{gather}
	G_{\mu\nu} = \eta_{AB} e^A_\mu e^B_\nu + \lambda^a_\mu \lambda^a_\nu , \qquad G_{\mu i} = \kappa^a_i \lambda^a_\mu , \qquad G_{ij} = \kappa^a_i \kappa^a_j ,
	\label{NATD:construct_G}
\end{gather}
where $\eta_{AB}$ is the seven-dimensional Minkowski metric. According to the definitions given in section \ref{secNATD} and using the frame fields in eq. (\ref{frames1}) we have:
\begin{gather}\label{quantities}
	\kappa^a_i = \sqrt{2}
	\begin{pmatrix}
		\lambda_2(r) & 0 & 0  \\
		0 & \lambda_2(r) & 0 \\
		0 & 0 & \lambda(r)
	\end{pmatrix} , \qquad 
	\lambda^a_\mu = \lambda(r) \cos \theta_1
	\begin{pmatrix}
		 0 & 0  \\
		 0 & 0  \\
		 0 & 1
	\end{pmatrix} .
\end{gather}
The outcome of the NATD procedure\footnote{Note that we will generally mark quantities in the T-dual background with carets.} is an $SU(2)_1\times U(1)_{\psi}$ dual background with NS fields given by
\begin{align}
	d\hat s^2 &= dy^{2}_{1,3}+ \kappa^{-1}dr^2+\lambda_1^2(r) (\sigma_{\hat 1}^2 + \sigma_{\hat 2}^2) + \frac{\lambda_2^2(r) \lambda^2(r)}{\Delta} x_1^2 \sigma_{\hat 3}^2 \nonumber\\ 
								   & \quad + \frac{1}{\Delta} \left[ (x_1^2 + \lambda^2(r) \lambda_2^2(r)) \, dx_1^2 + (x_2^2 + \lambda_2^4(r)) \, dx_2^2 + 2x_1 x_2 \, dx_1 \, dx_2 \right] , \label{dsdualresolved}\\
	\hat B &= -\frac{\lambda^2(r)}{\Delta} \left[ x_1 x_2 \, dx_1 + (x_2^2 + \lambda_2^4(r)) \, dx_2 \right] \wedge \sigma_{\hat 3} , \label{b1}\\
		e^{-2\hat\Phi} &= \frac{8}{g_s^2} \Delta \label{dil1} ,
\end{align}
where
\begin{align}
	 \sigma_{\hat 3} &\equiv d\psi + \cos \theta_1 \, d\phi_1 \label{sigma3},\\
	\Delta &\equiv \lambda_2^2(r) x_1^2 + \lambda^2(r) \left[ x_2^2 + \lambda_2^4(r) \right] \label{delta1} ,
\end{align}
Note the appearance of $\lambda_2$ in the quantity $\Delta$ and not $\lambda_1$. The above asymmetry between these two quantities has important physical consequences.
In the present case, we see that in the IR the dilaton is bounded away from $x_1\sim 0$. 
If we had chosen to dualize with respect to the $SU(2)_1$ isometry defined by  $(\theta_1,\phi_1,\psi)$, which involves the $S^3$ whose size shrinks to zero at the tip of the conifold, the NATD solution would still be given by the background in eqs. (\ref{dsdualresolved})-(\ref{dil1}) up to the trivial replacements $\lambda_2(r)\leftrightarrow \lambda_1(r)$, $\theta_2\leftrightarrow \theta_1$ and $\phi_2\leftrightarrow\phi_1$.
In this case the IR limit contains a dilaton that blows up.

The dual geometry is typically non-trivially related to the original one, which we see here also, but in contrast to the abelian T-duality we have no means to extract global topological information: we have only information at local patches of the manifold.
In the abelian case we could impose coordinate bounds on the dual coordinates by the requirement that the gauging procedure was applicable to worldsheets of arbitrary genus \cite{hep-th/9110053}.
However, for the non-abelian case it is a long standing problem on how to generalize the procedure beyond spherical worldsheets \cite{hep-th/9309039}, and we usually fall to physical constraints coming from the associated CFTs in order to extract global information \cite{Lozano:2013oma, Lozano:2014ata}, and we shall also do so here in order to extract coordinate bounds in section \ref{subsec:page}.


\section{Adding regular D3-branes}\label{sec:regularD3}
The configuration of $N$ D3-branes smeared over the $S_{2}^2$ at the tip of the resolved conifold was studied by
Pando-Zayas and Tseytlin (PT) in \cite{hep-th/0010088}.  The supergravity solution was found to be:
\begin{align}
	ds^2=&h(r)^{-1/2} \, dy_{1,3}^2 + h(r)^{1/2} \, ds_{6}^2,\label{ansatz}\\
F_5=&(1+\star)dh(r)^{-1}\wedge dy^{0}\wedge dy^{1}\wedge dy^{2}\wedge dy^{3},\label{5form}\\
&\qquad\Phi=\textrm{const},\label{dil2}
\end{align}
where $ds_6^2$ is the resolved conifold metric of eq. (\ref{dsresolved}) and
\begin{gather}
h(r)=\frac{2L^4}{9a^2r^2}-\frac{2L^4}{81a^4}\ln \left(1+\frac{9a^2}{r^2}\right).
\end{gather}
In the UV $(r\rightarrow \infty)$ this solution goes over to the KW solution of D3-branes at the conifold singularity \cite{Klebanov:1998hh}, but is singular in the IR as the warping factor behaves like
\begin{gather}
	h(r\rightarrow 0) \sim \frac{1}{r^2}.
\end{gather}
This singular behaviour is due to the smearing of the D3-branes and as a consequence the D3 Page charge blows up as well. This singularity can be avoided by localizing the D3-branes on the finite $S^2_2$ at the tip as was shown in \cite{Klebanov:2000hb}.

Let us proceed in T-dualizing this solution. Because of the smearing of the D3-branes, the metric in eq. (\ref{ansatz}) is still $SU(2)_1 \times SU(2)_2 \times U(1)_{\psi}$ invariant. Thus, as before, we dualize this background along the $SU(2)_2$ isometry of the internal space.
We proceed first to dualize the NS sector. We choose frame fields for the metric in eq. (\ref{ansatz}) as:
\begin{align}
	\tilde e^{y^\mu} &= h^{-\frac{1}{4}} \, dy^\mu,  \quad
	\tilde e^r = h^{\frac{1}{4}} \kappa^{-\frac{1}{2}}(r) \, dr , \quad
	e^{\theta_1 , \phi_1} = \lambda_1(r) \sigma_{\hat 1, \hat 2} ,\nonumber \\
	&\qquad e^{1,2} = \lambda_2(r) \sigma_{1,2} , \quad
	e^3 = \lambda(r) (\sigma_3 + \cos \theta_2 \, d\phi_2) ,
	\label{framesd3}
\end{align}
where the warping factor has been absorbed into the redefined $\lambda$'s
\begin{gather}
	\lambda_1^2(r) \equiv h^{\frac{1}{2}} r^2/6, \quad
	\lambda_2^2(r) \equiv h^{\frac{1}{2}} \left(r^2 + 6 a^2\right)/6 , \quad
	\lambda^2(r) \equiv h^{\frac{1}{2}} \kappa(r) \, r^2/9 .
	\label{redlambdah}
\end{gather}
The background written in this form resembles the form of the one studied in section \ref{sec1}, but it should be remembered that the $\lambda$'s of eq. (\ref{redlambdah}) are not the same as with the pure NS solution $\lambda$'s of eq. (\ref{lambdas}).  We now gauge fix in such a way that all three Lagrange multipliers $v_i$ will be promoted to coordinates in the dual geometry,  i.e. making the choice $g=\mathbb{I}$. We obtain a solution of Type IIA supergravity  with NS fields given by
\begin{align}	
	d\hat s^2 &= h(r)^{-1/2} \, dy_{1,3}^2 + h(r)^{1/2} \, \kappa(r)^{-1} dr^2 + \lambda_1^2(r) (\sigma_{\hat 1}^2 + \sigma_{\hat 2}^2) + \sum_{a=1}^3\hat{e}^{a}\hat{e}^{a},  \label{dsresolvedd3} \\
	\hat B &= -\frac{1}{\Delta}\Big[x_3\lambda^2(r) dx_1\wedge dx_2 +\lambda_2^2(r)\left(x_1 dx_2-x_2 dx_1\right)\wedge dx_3  \nonumber\\
	&\qquad+ \lambda(r)\cos\theta_1\left(x_1 x_3 dx_1+x_2 x_3 dx_2+(x_3^2+\lambda_2^4(r)\right)dx_3)\wedge d\phi_1\Big]
	\label{b2d3} , \\
	e^{-2\hat\Phi} &= \frac{8}{g_s^2} \Delta \label{dil3},
\end{align}
where 
\begin{align}
\hat{e}^{1}&=-\frac{\lambda_2(r)}{\Delta}\left((x_1^2+\lambda^2(r)\lambda_2^2(r))dx_1+(x_1 x_2+x_3\lambda^2(r))dx_2+(x_1x_3-x_2\lambda^2(r))\overline{\sigma}_{\hat{3}}\right), \nonumber \\
\hat{e}^{2}&=-\frac{\lambda_2(r)}{\Delta}\left((x_2^2+\lambda^2(r)\lambda_2^2(r))dx_1+(x_1 x_2-x_3\lambda^2(r))dx_2+(x_2x_3+x_1\lambda^2(r))\overline{\sigma}_{\hat{3}}\right), \nonumber \\
\hat{e}^3&=-\frac{\lambda(r)}{\Delta}\left( (x_1x_3+x_2\lambda^2(r))dx_1+(x_2x_3-x_1\lambda^2(r))dx_2+\overline{\sigma}_{\hat{3}}\right)+\lambda(r)\cos\theta_1 d\phi_1,
\end{align}
and 
\begin{align}
\overline{\sigma}_{\hat{3}}&\equiv dx_3+\lambda^2(r)\cos\theta_1 d\phi_1, \nonumber\\
\Delta&\equiv \lambda^2(r)x_3^2+\lambda_2^2(r)(x_1^2+x_2^2+\lambda_2^2(r)).
\end{align}
One can easily see that the T-dual background metric of eq. (\ref{dsresolvedd3}) has inherited the $1/r^2$ singular behavior in the IR.
 However, no more singularities will appear\footnote{Up to the bolt singularity that is avoided by reducing the range of $\psi$ to $2\pi$}.
In the UV, the background in eqs. (\ref{dsresolvedd3})-(\ref{dil3}) asymptotes to the NATD KW background \cite{Itsios:2012zv, Itsios:2013wd} since
\begin{gather}
	h(r \rightarrow \infty) = \frac{1}{r^4}, \quad
	\lambda_1^2(r \rightarrow \infty) = \lambda_2^2(r \rightarrow \infty) = 1/6, \quad
	\lambda^2(r \rightarrow \infty) = 1/9.
\end{gather}
The behaviour in this limit is independent of the $SU(2)$ over which we decide to dualize. This is to be expected because far away from the tip the two $S^2$ submanifolds of the resolved conifold are indistinguishable.
 Again, if we had chosen to dualize 
about the $SU(2)_1$ isometry, the T-dual NS fields would be given  as in eq. (\ref{dsresolvedd3})-(\ref{dil3}) up to trivial replacements. In this case we will have a dilaton which blows up in the IR\footnote{Due to the presence of the RR fields one can try to up-lift this solution to eleven dimensions in order to protect the geometry from being strongly coupled.}.

Let us now turn  to dualizing  the RR sector. The 5-form flux in eq. (\ref{5form}) is explicitly
\begin{gather}
F_5= p(r) \left[ e^{y^0} \wedge  e^{y^1} \wedge  e^{y^2} \wedge  e^{y^3} \wedge  e^r  -  e^{\theta_1} \wedge e^{\phi_1} \wedge e^1 \wedge e^2 \wedge e^3 \right] ,
\end{gather}
where
\begin{gather}
	p(r)= - h(r)^{-5/4} \kappa(r)^{1/2} \frac{dh(r)}{dr} .
\end{gather}
The T-dual fluxes\footnote{Notice that the orientation we have used here for the Maurer-Cartan forms of eq. (\ref{eulerforms}) differs from the ones used in \cite{Itsios:2013wd} via the interchange $\sigma_1 \leftrightarrow \sigma_2$, and this will induce an extra minus sign in the RR fluxes calculated here.} are found to be
\begin{align}
\hat F_2 &= 2\sqrt{2} \, p(r) \, \lambda_1^2 \, \lambda_2^2 \, \lambda \, \sigma_{\hat 1} \wedge \sigma_{\hat 2} ,\label{rrd3} \\
	\hat F_4 &=-2\sqrt{2} \, p(r)\Big(\lambda (r)x_3 e^{\theta_1}\wedge e^{\phi_1}\wedge \hat{e}^{1}\wedge \hat{e}^2 \nonumber \\
&\qquad\qquad\qquad\quad +\lambda_2(r)\left(x_1 e^{\theta_1}\wedge e^{\phi_1}\wedge \hat{e}^2\wedge \hat{e}^3-x_2 e^{\theta_1}\wedge e^{\phi_1}\wedge \hat{e}^1\wedge \hat{e}^3\right)\Big)\nonumber
\end{align}
which support the NS T-dual background of eqs. (\ref{dsresolvedd3})-(\ref{dil3}). 
We have written down only the lower order RR fluxes since the higher order ones can be obtained from $F_p=(-1)^{p/2}\star F_{10-p}$.


\section{Adding fractional and regular D3-branes}\label{sec:reg+fracD3}
A generalization of the KT solution, where fractional and regular D3-branes were placed on the conifold \cite{Klebanov:2000nc}, was considered in \cite{hep-th/0010088} by placing this configuration on the resolved conifold.
This solution was obtained by starting with an ansatz similar to the one in eq. (\ref{ansatz}), but now we find an $SU(2)_1\times SU(2)_2\times U(1)_{\psi}$ symmetric solution with the warping factor satisfying
\begin{gather}
	\frac{dh}{dr} = -108 r^{-3} (r^2 + 9 a^2)^{-1} K(r)\label{warpd3d5},
\end{gather}
where $K(r)$ is defined in eq. (\ref{Kdefn}) below, and is related to the number of regular and fractional D3-branes.
This background is also supported by a self-dual 5-form and non-trivial NS and RR 3-form fluxes given by\footnote{Note that although some papers have declared a sign error in the RR 3-form flux, we find that the expression noted here, which is exactly the one presented in \cite{hep-th/0010088}, solves the Bianchi and flux equations, and is indeed a solution to the Type IIB supergravity equations of motion. We believe, therefore, that there is no sign error and that the fluxes both here and in \cite{hep-th/0010088} are correct.}
\begin{gather}
	H_3 = dr \wedge \left( \frac{d(\lambda_1(r)^{-2} f_1)}{dr} \, e^{\theta_1} \wedge e^{\phi_1} - \frac{d(\lambda_2(r)^{-2} f_2)}{dr} \, e^{1} \wedge e^{2}\right) ,\\
	F_3	= P \, \lambda(r)^{-1} \, e^3 \wedge ( \lambda_2(r)^{-2} \, e^{1} \wedge e^{2} - \lambda_1(r)^{-2} \, e^{\theta_1} \wedge e^{\phi_1} ) , \\
	F_5 = (1 + \star) \, K(r) \lambda(r)^{-1} \lambda_1(r)^{-2} \lambda_2(r)^{-2} \, e^3 \wedge e^1 \wedge e^2 \wedge e^{\theta_1} \wedge e^{\phi_1},
\end{gather}
where 
\begin{align}
	f_1(r) &= \frac{3}{2} P e^\Phi \ln  \left( r^2 +9 a^2 \right),\label{f1} \\
	f_2(r) &= \frac{1}{6} P e^\Phi \left[ \frac{36 a^2}{r^2} - \ln\left( r^{16} (r^2 + 9 a^2)\right)\right],\label{f2} \\
	K(r) &= Q + P(f_1 - f_2),\label{Kdefn}
\end{align}
where $P$ and $Q$ are proportional to the number $N$ and $M$ of fractional and regular D3-branes respectively, $\Phi$ is a constant dilaton field
and the frame fields and $\lambda$'s are those defined in eq. (\ref{framesd3}) and (\ref{redlambdah}) but with the warping factor in them satisfying eq. (\ref{warpd3d5}).
In the UV this solution reduces to the Klebanov-Tseytlin (KT) solution as $h(r)\sim \ln r$ whilst in the IR the solution has a naked singularity.

As in the previous sections, we shall proceed to dualize this background along the $SU(2)_2$ isometry of the internal space. 
The quantities $\kappa_i^{a}$ and $\lambda_{\mu}^{a}$ are unchanged from their form in eq. (\ref{quantities}) except, again, with $\lambda_2(r)$
and $\lambda(r)$ defined in  eq. (\ref{redlambdah}) but with the warping factor in them satisfying eq. (\ref{warpd3d5}). However, we now have a non-zero NS B-field, which we deconstruct following eq. (\ref{Bdeconstruction}) as:
\begin{gather}
B_{\mu \nu} = \frac{1}{2} f_1(r) \, \sin \theta_1
	\begin{pmatrix}
		  0  & 1 \\
		  -1 & 0
	\end{pmatrix} , \quad
	B_{ij} = f_2(r) \, \lambda_2^{-2}
	\begin{pmatrix}
		 0 & -1 & 0  \\
		 1 &  0 & 0  \\
		 0 &  0 & 0
	\end{pmatrix} .
\end{gather}
We shall use the same gauge fixing as in section \ref{sec1}. The result of the dualization is a solution of massive Type IIA supergravity with NS fields 
\begin{align}
	d\hat s^2 &= h(r)^{-1/2} \, dy_{1,3}^2 + h(r)^{1/2} \, \kappa(r)^{-1} dr^2 + \lambda_1^2(r) \left(\sigma_{\hat 1}^2 + \sigma_{\hat 2}^2\right)   +  \frac{\lambda_2^2(r) \lambda^2(r)}{\Delta} x_1^2 \sigma_{\hat 3}^2 \label{dualm6}  \\
			  &\qquad+ \frac{1}{\Delta} \Big[ (x_1^2 + \lambda^2(r) \lambda_2^2(r)) \, dx_1^2+ \left[(x_2 - f_2)^2 + \lambda_2^4(r)\right] \, dx_2^2 + 2x_1 (x_2 - f_2) \, dx_1 \, dx_2 \Big] ,  \nonumber \\
			  \hat B &= f_1(r) \, \sigma_{\hat 1} \wedge \sigma_{\hat 2} - \frac{1}{\Delta} \Big\{ \lambda^2(r) \, x_1 \left(x_2-f_2\right) \, dx_1 \wedge \sigma_{\hat 3} \\
&\qquad+ dx_2 \wedge \left[ x_1^2 \lambda_2^2 \, d\psi - \lambda^2 \left( \left(x_2-f_2\right)^2 + \lambda_2^4 \right) \cos\theta_1 \,d\phi_1 \right] \Big\} , \nonumber\\
	&\qquad\qquad\qquad \qquad\qquad \qquad\qquad e^{-2\hat\Phi} = 8\Delta \label{dil6},
				\end{align}
where
\begin{equation}
\begin{split}
	\sigma_{\hat 3} &\equiv d\psi + \cos \theta_1 \, d\phi_1 , \\
	\Delta &\equiv \lambda_2^2(r) \, x_1^2 + \lambda^2(r) \left[(x_2 - f_2)^2 + \lambda_2^4(r)\right] . \label{regfrac:delta} \\
	\end{split}
\end{equation}
Note that the presence of the fractional D3-branes has induced a shift in the T-dual coordinate $x_2$ by $x_2 \rightarrow x_2 - f_2(r)$, but leaves $dx_2$ unchanged.
In the IR the solution has a naked singularity and we expect that the arguments of \cite{hep-th/0010088} in avoiding the singularity in the original PT solution via an enhancon-type mechanism \cite{Johnson:1999qt} will also apply here.
Note also that in the UV this solution reduces to the T-dual of the Klebanov-Tseytlin solution\footnote{This is more clear in the coordinate system in which the T-dual coordinates are $(x_2,x_3,\psi)$. Note also that the definitions here differ from the ones in \cite{Itsios:2013wd} as $T\rightarrow \frac{T}{6\sqrt{2}}, ~P\rightarrow \frac{ P}{18\sqrt{2}},~K\rightarrow \frac{K}{108}$.} \cite{Itsios:2013wd}.

The RR fluxes supporting this T-dual solution are given by
\begin{equation}
\begin{split}
	\frac{1}{2\sqrt{2}} \hat{F}_0 &=  P , \\
	\frac{1}{2\sqrt{2}} \hat{F}_2 &= - \frac{P \, x_1}{\Delta} \left[ \lambda^2 \left(x_2 - f_2\right) dx_1  -  \lambda_2^2 \, x_1 dx_2 \right] \wedge \sigma_{\hat 3}  - \left[P \left(x_2 - f_2\right) - K \right] \sigma_{\hat 1} \wedge \sigma_{\hat 2} , \\
	\frac{1}{2\sqrt{2}} \hat{F}_4 &= \frac{x_1}{\Delta} \sigma_{\hat 1} \wedge \sigma_{\hat 2} \wedge \sigma_{\hat 3} \wedge \Big\{ \left[ K \lambda^2 \left(x_2 - f_2\right) + P \lambda_2^2 \left( x_1^2 + \lambda^2 \lambda_2^2 \right) \right] dx_1   \\
		&\qquad\qquad\qquad\qquad\qquad\qquad\qquad\qquad\qquad+ \lambda_2^2 x_1 \left[ P \left(x_2 - f_2\right) - K \right] dx_2 \Big\} ,
\end{split}
\label{fluxes_lower}
\end{equation}
with the higher-order forms given by
\begin{equation}
\begin{split}
	\frac{1}{2\sqrt{2}} \hat{F}_6 &= - \frac{\textrm{Vol}_{5}}{\lambda \lambda_1^2 \lambda_2^2} \wedge \left\{ K x_1 dx_1 + \left[ K \left(x_2 - f_2\right) + P \lambda_2^4 \right] dx_2 \right\} , \\
	\frac{1}{2\sqrt{2}} \hat{F}_8 &= - \textrm{Vol}_{5} \wedge \Big\{ \frac{\lambda_2^2 \lambda \, x_1}{\lambda_1^2 \Delta} \left[ P \left(x_2 - f_2\right) - K \right] dx_1 \wedge dx_2 \wedge \sigma_{\hat 3}  \\
		&\quad- \frac{P}{\lambda_2^2 \lambda} \left[ \lambda_1^2 \, x_1 \, dx_1 + \lambda_1^2 \left(x_2 - f_2\right) dx_2 \right] \wedge \sigma_{\hat 1} \wedge \sigma_{\hat 2} \Big\} , \\
	\frac{1}{2\sqrt{2}} \hat{F}_{10} &= - \frac{ P \, x_1 \, \lambda_1^2 \, \lambda_2^2 \, \lambda}{\Delta} \textrm{Vol}_{5} \wedge dx_1 \wedge dx_2 \wedge  \sigma_{\hat 1} \wedge \sigma_{\hat 2} \wedge \sigma_{\hat 3},
\end{split}
\label{fluxes_higher}
\end{equation}
where $\textrm{Vol}_5 = h^{-3/4} \kappa^{-1/2} dy^0 \wedge dy^1 \wedge dy^2 \wedge dy^3 \wedge dr$.
A straightforward computation shows that the dual fluxes in eq. (\ref{fluxes_lower}) are related to the ones in eq. (\ref{fluxes_higher}) by $F_p=(-1)^{p/2}\star F_{10-p}$.


\subsection{Comments on the G-structure of the solution}\label{Gstructure}
We now turn to the discussion of the supersymmetry of this solution. As we stated in the introduction, it was shown in \cite{Cvetic:2000db} that the PT solution prior to dualisation
is non-supersymmetric. Here we shall prove this fact in a slightly different way, as this will allow us to analyse the G-structure of the backgrounds prior and after the duality transformation as well as the proper for the case of section \ref{sec:regularD3}. One could study the supersymmetry of these solutions by just analyzing the Killing spinors of the solutions to see if they
are independent of the $SU(2)$ directions along which we performed the duality. If so then supersymmetry will be preserved under the duality \cite{Kelekci:2014ima}. 

Let us suppose that the PT solution discussed in section \ref{sec:reg+fracD3} defines an $\mathcal{N}=1$ supersymmetric vacuum. This means that one would be able to define an $SU(3)$ structure in this background.
According to the frame fields for this solution, with the $\lambda$'s defined in eq. (\ref{framesd3}) and (\ref{redlambdah}) and the warping factor in them satisfying eq. (\ref{warpd3d5}),
 the structure that hypothetically characterizes this solution would be given by
\begin{gather}
J= e^{r3}+e^{\varphi_1\theta_1}+e^{12},\\\Omega_{hol}=(e^r+ i e^3)\wedge(e^{\varphi_1}+ie^{\theta_1} ) \wedge(e^{1}+ie^{2} ),
\end{gather}
from which we can construct two pure spinors
\begin{equation}
\Psi_+ = \frac{1}{8}e^{i \theta_+} e^{A} e^{-i J},~~~~\Psi_- = -i\frac{1}{8} e^{A}\Omega_{hol}\label{pspinors},
\end{equation}
where $\theta_{\pm}$ are arbitrary phases. Using the background metric ansatz in eq. (\ref{ansatz}) for this solution, written in the form 
\begin{gather}
ds^2=e^{A}dy_{1,3}^2+ds^{2}(\mathcal{M}^6),
\end{gather}
we find also that $e^{A}=\frac{1}{\sqrt{h(r)}}$. 
In order to have a solution with $\mathcal{N}=1$ supersymmetry, the above pure spinors  must satisfy the following differential conditions \cite{Grana:2005sn}:
\begin{align}
&\big(d-H\big) e^{2A-\Phi} \Psi_-=0,\\
&\big(d-H\big) e^{2A-\Phi} \Psi_+-e^{2A-\Phi} dA\wedge \bar{\Psi}_+-i\frac{e^{3A}}{8} \star_6 \tilde{F}=0,
\end{align}
where $\tilde{F}=F_5+(1-\star) F_3$. By direct computation one can easily verify that the first of these equations is automatically satisfied, whereas the second differs from zero by a term $\sim 27 a^2 P$, from which we conclude that
this solution does not represent an $\mathcal{N}=1$ vacuum. Note that the equations are satisfied if either $a=0$ or $P=0$. The former case corresponds to the KT solution whilst the latter corresponds to the solution studied in section \ref{sec:regularD3}. In both cases the pure spinors in eq. (\ref{pspinors}) define an $SU(3)$ structure 
with $\Omega_{hol}$ and $J$ satisfying 
\begin{gather}
J\wedge \Omega_{hol}=0, \quad J\wedge J\wedge J=\frac{3i}{4}\Omega_{hol}\wedge \bar{\Omega}_{hol} ,
\end{gather}
and 
\begin{gather}
\theta_{+}=\frac{\pi}{2}, \quad \theta_{-}=0 .
\end{gather}
Let us now turn our attention to the structure of the T-dual background of eqs. (\ref{dualm6})-(\ref{dil6}) and (\ref{fluxes_lower}).
Let us define a new set of internal frame fields via $\tilde{e}=R\,\hat{e}$, where the rotation matrix $R$ coincides with the one presented in \cite{Barranco:2013fza} up to the replacement\footnote{Again, this is due to the orientation we have used for the Maurer-Cartan forms differing by the reflection $\sigma_1 \leftrightarrow \sigma_2$ when we T-dualized the background.} $\zeta^1\leftrightarrow \zeta^2$,  with 
\begin{gather}
\zeta^1=\frac{x_1}{\lambda(r)\lambda_2(r)},\quad  \zeta^2=\frac{x_2}{\lambda(r)\lambda_2(r)},\quad  \zeta^3=\frac{x_3}{\lambda_2^2(r)}.
\end{gather}
Explicitly, the new frame fields are
\begin{align}
\tilde{e}_r=&\frac{\lambda(r)^{-1}\lambda_{2}(r)^{-2}}{\sqrt{\Delta}\sqrt{\kappa}}\left(h(r)^{1/4}\lambda(r)\lambda_{2}(r)^{2}dr-\sqrt{\kappa}(x_1 dx_1+x_2 dx_2+x_3 dx_3)\right),\nonumber\\
&\qquad \qquad \tilde{e}_{\theta_1}=\lambda_1(r) d\theta_1,\qquad \tilde{e}_{\phi_1}=\lambda_1(r)\sin\theta_1 d\phi_1,\nonumber\\
\tilde{e}_{1}=&-\frac{\lambda(r)^{-1}\lambda_{2}(r)^{-1}}{\sqrt{\Delta}\sqrt{\kappa}}\left(x_2 h(r)^{1/4}dr+\sqrt{\kappa}\lambda(r) (dx_2+x_1 \cos\theta_1 d\phi_1) \right),\\
\tilde{e}_{2}=&-\frac{\lambda(r)^{-1}\lambda_{2}(r)^{-1}}{\sqrt{\Delta}\sqrt{\kappa}}\left(x_1 h(r)^{1/4}dr+\sqrt{\kappa}\lambda(r) (dx_1-x_2 \cos\theta_1 d\phi_1) \right),\nonumber\\
\tilde{e_3}=&\frac{\lambda(r)^{-1}\lambda_{2}(r)^{-2}}{\sqrt{\Delta}\sqrt{\kappa}}\left(-x_3 h(r)^{1/4}dr-\sqrt{\kappa}\lambda_2(r)^2  dx_3  \right). \nonumber
\end{align}
It is then straightforward to check that the T-dual background of  eqs. (\ref{dualm6})-(\ref{dil6}) and (\ref{fluxes_lower}) support two pure spinors
\begin{gather}
\Phi_{+}=-i\frac{e^{A}}{8}e^{i\theta_{+}}e^{i v\wedge w}\omega, \quad \Phi_{-}=i\frac{e^{A}}{8}e^{i\theta_{-}}( v+i w)\wedge e^{-ij },
\end{gather}
where the $SU(2)$ 2-forms $\omega, j$,  2-forms $v, w$ and  phases are given by 
\begin{gather}
j=\tilde{e}^{r3}+\tilde{e}^{\phi_1\theta_1}+\tilde{e}^{21}, \quad  \omega= (\tilde{e}^{\phi_1}+i\tilde{e}^{\theta_1})\wedge\left(\tilde{e}^2+i\tilde{e}^1\right)\nonumber , \\
 z=v+iw=\tilde{e}^3+i\tilde{e}^r,\\
\tilde{\theta}_{+}=\frac{\pi}{2},\quad \tilde{\theta}_{-}=0,\nonumber
\end{gather}
which defines an orthogonal $SU(2)$ structure. In the UV, this structure flows to the orthogonal $SU(2)$ structure of the KW solution presented in \cite{Barranco:2013fza}.
The IR limit does not have an $SU(2)$ structure, as expected.

\section{Features of the NATD geometry and the dual field theory}
\label{sec:probes}
In this section we shall study some quantities that define features of the PT background and the dual field theory associated with it prior to and after dualization.
First we consider the D-brane Page charges to see how these quantities get mapped under NATD.
We shall also see that the definition of a gauge coupling in the dual field theory may lead us to impose bounds on the T-dual coordinates.
In order to gain an insight of how field theory features behave under the NATD, we shall finally study the central charge.

\subsection{Page charges}
\label{subsec:page}
In order to examine the content of the dual background, let us calculate the Page charges of D-branes before and after T-dualization.
We shall start with the undualized solution.
The Page charge of D3-branes is given by
\begin{equation}
	Q_{D3} = \int F_5 - B_2 \wedge F_3 + \frac{1}{2} B_2 \wedge B_2 \wedge F_1 ,
\end{equation}
which is explicitly
\begin{equation}
	Q_{D3} = \int Q \sin\theta_1 \sin\theta_2 \, d\theta_1 \wedge d\phi_1 \wedge d\theta_2 \wedge d\phi_2 \wedge d\psi = (4\pi)^3 Q ,
\end{equation}
which, as expected, is quantized by the number of regular D3-branes on the background.

The Page charge of D5-branes is 
\begin{equation}
	Q_{D5} = \int F_3 - B_2 \wedge F_1. 
\end{equation}
We find two 3-cycles in the geometry defined by the submanifolds
\begin{gather}
\Sigma_1=(\theta_1,\phi_1,\psi),\quad  \Sigma_2=(\theta_2,\phi_2,\psi),
\end{gather}
on which we can compute the D5-brane charge to be
\begin{equation}
	Q_{D5} = -P \int_{\Sigma_1} \sin\theta_1 \, d\theta_1 \wedge d\phi_1 \wedge d\psi = - (4\pi)^2 P ,
\end{equation}
and
\begin{equation}
	Q_{D5} = P \int _{\Sigma_2} \sin\theta_2 \, d\theta_2 \wedge d\phi_2 \wedge d\psi = (4\pi)^2 P .
\end{equation}
Therefore, we see that the D5-brane Page charges are quantized by the number of fractional D3-branes, again, as would be expected.

Now we turn to the computation of charges after NATD.
Following the discussion in \cite{Itsios:2013wd} we find that the Page charge of D8-branes become
\begin{equation}
	Q_{D8} = \int \hat{F}_0 = 2\sqrt{2} P ,
\end{equation}
and it is clear that the quantization of the Romans mass is given by the number of fractional D3-branes prior to the dualization.

The D6 Page charge is given by
\begin{equation}
	Q_{D6} = \int \hat{F}_2 - \hat{F}_0 \hat{B}_2 ,
\end{equation}
and we find that
\begin{equation}
	\hat{F}_2 - \hat{F}_0 \hat{B}_2 = 2\sqrt{2} \left[ P \cos\theta_1 \, dx_2 \wedge d\phi_1 - \left(P \, x_2 + Q \right) \sin\theta_1 \, d\theta_1 \wedge d\phi_1  \right] .
\end{equation}
It is expected that the NATD procedure will transform D3-brane Page charge into D6-brane Page charge, which is only consistent in our case if we choose two-cycles over $(\theta_1, \phi_1)$ quantized at the points $x_2 \in \mathbb{Z}$. Then, we find that the D6-branes seem to wrap a 2-sphere at integer points along the $x_2$ coordinate and the charge itself is a linear combination of the number of regular and fractional D3-branes before the NATD procedure. Perhaps the alternative viewpoint is that the $x_2$ coordinate in some way winds around a 2-sphere and that we are re-counting charge as we cycle around the $x_2$ coordinate.

The Page charge of D4-branes is given by
\begin{align}
	Q_{D4} &= \int \hat{F}_4 - \hat{B}_2 \wedge \hat{F}_2 + \frac{1}{2} \hat{F}_0 \hat{B}_2 \wedge \hat{B}_2  \nonumber \\
		   &=  -2\sqrt{2} P \int \sin\theta_1 \, dx_2 \wedge d\theta_1 \wedge d\phi_1 \wedge d\psi .
\end{align}
However, since the RR fields given in eq. (\ref{fluxes_lower}), satisfy the Bianchi equations
\begin{gather}
	d\hat{F}_2 = \hat{H}_3 \hat{F}_0 , \qquad
	d\hat{F}_4 = \hat{H}_3 \wedge \hat{F}_2 ,
\end{gather}
then the Page charge of D4-branes should be identically zero. In fact, we can make it so by gauging the $\hat{B}_2$ field:
\begin{equation}
	\hat{B}_2 \rightarrow \hat{B}_2 + \mu_2 , \qquad\textrm{where}\quad d\mu_2 = 0 .
\end{equation}
The choice of $\mu_2$ is not unique; for example we can make the choice
\begin{align}
	\mu_2 &= g(x_1, \theta_1) \, dx_1 \wedge d\theta_1 - x_1 dx_1 \wedge d\phi_1 + \cos\theta_1 dx_2 \wedge d\phi_1  \nonumber \\
		  &\qquad- x_2 \sin\theta_1 d\theta_1 \wedge d\phi_1 + \sin\theta_1 d\theta_1 \wedge d\psi ,
\end{align}
where $g(x_1, \theta_1)$ is an arbitrary function.

The D2-brane Page charge is
\begin{align}
	Q_{D2} &= \int \hat{F}_6 - \hat{B}_2 \wedge \hat{F}_4 + \frac{1}{2} \hat{B}_2^2 \wedge \hat{F}_2 - \frac{1}{6} \hat{F}_0 \hat{B}_2^3, \nonumber \\
					   &=\int - \frac{2\sqrt{2}}{\lambda \lambda_1^2 \lambda_2^2 } \textrm{Vol}_{5} \wedge \left( K(r) \,x_1\, dx_1 + \left[\left(x_2 - f_2\right) K(r) + P \lambda_2^4 \right] dx_2 \right) ,
\end{align}
for which no 6-cycle will quantize the charge since all the terms run with the holographic radial coordinate.

Thus, we conclude that the D3 and D5-brane charges before dualization have transformed into D6 and D8-brane charge. To quantize the D6-brane Page charge we have to impose either a quantization of the two-cycle with respect to the dual coordinate $x_2$ or else some type of topological winding of the $x_2$ coordinate around a 2-sphere, but otherwise the D3 charge would be destroyed by the NATD procedure.
\subsection{The central charge}

On the field theory side there are some quantities which define and characterize the theory.
One of these quantities is the central charge, which is a measure of the number of degrees of freedom in the field theory.
Here we will be interested in computing this quantity for the NATD PT solution presented in section \ref{sec:reg+fracD3} and comparing it with the calculation prior to dualization.
We will follow the technique developed in \cite{Klebanov:2007ws} and then used in \cite{Itsios:2013wd} for backgrounds obtained as the result of NATD.
Consider a ten-dimensional string-frame metric of the form
\begin{gather}
ds^2=\alpha dy_{1,3}^2+\alpha\beta dr^2+ds^{2}(\mathcal{M}^6),
\end{gather}
for which the central charge is defined by
\begin{gather}
c=27\beta^{3/2}\frac{H^{7/2}}{(H')^3},\qquad H=V_{int}\alpha^3,
\end{gather}
where $V_{int}$ is the volume of $\mathcal{M}^6$. 
The volume $V_{int}$ before dualization is well defined: the coordinates have ranges $0 \le \theta_1 \le \pi$, $0 \le \phi_1 < 2\pi$, $0 \le \theta_2 \le \pi$, $0 \le \phi_2 < 2\pi$, $0 \le \psi < 4\pi$.
However it is not known how to determine the coordinate ranges after non-abelian T-dualization, and following \cite{Itsios:2013wd} and \cite{Lozano:2014ata} we will examine associated field theory quantities. First we rewrite the NATD results of eqs. (\ref{dualm6})-(\ref{regfrac:delta}) with the gauge fixing choice $g=\mathbb{I}$, then  we perform a coordinate transformation to spherical polar coordinates as
\begin{equation}
	\frac{v_1}{2} \equiv \rho \sin\chi \cos\xi, \qquad \frac{v_2}{2} \equiv \rho \sin\chi \sin\xi, \qquad \frac{v_3}{2} \equiv \rho \cos \chi, 
	\label{transf}
\end{equation}
where $0 \le \xi < 2\pi$, $0 \le \chi \le \pi$ and $\rho$ has a nominal range $\rho > 0$.
We then find that the dual NS B-field takes the form:
\begin{align}
		\hat{B}_2   &= f_1(r) \, \sigma_{\hat 1} \wedge \sigma_{\hat 2}  + \frac{1}{\Delta} \Big\{  \rho \sin^2 \chi \left[ \lambda^2 \left(\rho \cos\chi - f_2\right) - \lambda_2^2 \, \rho \cos\chi \right] d\xi \wedge d\rho  \nonumber \\
				&\qquad+ \rho^2 \sin\chi \left[ \lambda^2 \cos\chi \left(\rho \cos\chi - f_2 \right) + \lambda_2^2 \, \rho \sin^2 \chi \right] d\xi \wedge d\chi  \nonumber \\
	   			&\qquad+ \lambda^2 \cos\theta_1 \left[ \rho f_2 \left(1+\cos^2 \chi\right) - \cos\chi \left(f_2^2 + \rho^2 +\lambda_2^4 \right) \right] d\rho \wedge d\phi_1  \\
				&\qquad+ \lambda^2 \rho \sin\chi \cos\theta_1 \left[ f_2 \left(\rho \cos\chi - f_2\right) - \lambda_2^4 \right] d\phi_1 \wedge d\chi \Big\} ,  \nonumber 
\end{align}
where
\begin{equation}
\begin{split}
	\sigma_{\hat 3} &\equiv d\xi + \cos \theta_1 \, d\phi_1 , \\
	\Delta &\equiv \lambda_2^2(r) \, \rho^2 \sin^2 \chi + \lambda^2(r) \left[(\rho \cos\chi - f_2)^2 + \lambda_2^4(r)\right] . 
	\end{split}
\end{equation}
On the manifold $\rho = \textrm{const}, \theta_1 = 0, \phi_1 = 2\pi-\xi$, the dual $\hat{B}_2$ form is
\begin{equation}
	\hat{B}_2 = \rho \sin\chi d\xi \wedge d\chi.
\end{equation}
If we examine the quantity
\begin{equation}
	b_0 = \frac{1}{4\pi^2} \int_{S^2} \hat{B}_2 ,
\end{equation}
which is related to the gauge coupling in the dual field theory, we then find
\begin{equation}
	b_0 = \frac{\rho}{4\pi^2} \int \sin\chi d\xi \wedge d\chi = \frac{\rho}{\pi}.
	\label{coupling}
\end{equation}
Since this is required to be bounded,  $0 < b_0 < 1$, then we conclude that the dual field theory is invariant under $\rho \rightarrow \rho + n\pi$, and that in a physical sense the NATD internal geometry is compact. 
A straightforward computation shows that the central charges before and after dualization are related by
\begin{gather}
	\hat{c}=\frac{2\rho_\textrm{max}^3}{3\pi}c,
	\label{ccharge}
\end{gather}
where $\rho_\textrm{max}$ is the upper bound of $\rho$.
With the above argument, we can explicitly evaluate eq. (\ref{ccharge}) to be
\begin{gather}
\hat{c}=\frac{2\pi^2}{3}c.
\end{gather}

The numerical factor between the central charges indicates how, and by how much, the NATD has influenced the theory, and as with other results of NATD we see that they are related by a constant of proportionality.

\section{Conclusions}\label{cremarks}
In this work we have studied new solutions of Type IIA and massive Type IIA supergravity obtained by performing a non-abelian T-duality along $SU(2)$ isometries of warped D3-brane solutions on the resolved conifold.
We began with the pure NS resolved conifold geometry, and after implementing the NATD we find a smooth geometry supported by a non-zero B-field and non-trivial dilaton that proved to be a Type IIA pure NS supergravity solution, which in the IR could acquire strongly coupled behaviour depending on the choice of $SU(2)$ to dualize over. When we dualize the $SU(2)$ isometry that contains the $S^2_2$ that has a finite size at the conifold tip, we find that the dilaton is regular as long as we stay away from the point $x_1 = 0$, whereas when we dualize over the $SU(2)$ isometry of the $S^3$ whose size shrinks to zero at the conical tip we have a dilaton which blows up.
Here an uplift to M-theory could not prevent the theory from being strongly coupled since there are no RR fluxes supporting the geometry.
In the UV we find that the details of the conifold resolution fall away and we get a smooth solution everywhere.

Next we added D3-branes to the resolved conifold geometry and we studied the effect of non-abelian T-duality in this background. As expected, after the duality transformation we obtained a solution of Type IIA supergravity that inherited the IR singular behaviour of the original solution, which was due to the smearing of the D3-branes at tip.
Therefore, one is led to conclude that the origin of the singularity in the NATD solution is due to the smearing of the D6 charge.
Far in the UV the solution asymptotes to the NATD of the KW solution.

A rather more interesting solution that we considered, and which generalizes the cases discussed above, is when we place, in addition to regular D3-branes, also fractional D3-branes at the tip of the resolved conifold.
We constructed the non-abelian T-dual of this solution and found that the dual background is a solution of massive Type IIA supergravity in which the Romans mass is quantized by the number of fractional D3-branes prior to the dualization.
By analyzing the Page charges of the backgrounds both before and after dualizing, we find that D3-brane charge is transformed into D6-brane charge, and the D5-brane Page charge contributes to both the D6-branes and D8-brane charge. In order to preserve charge across the NATD, we seemingly must quantize the $x_2$ dual coordinate, or else interpret it as a kind of winding over the residual $S^2$ of the conifold base.
We also studied the non-supersymmetric nature of the system of fractional and regular D3-branes by analyzing the structure of the background. We saw that in order to have an $\mathcal{N}=1$ solution we must set to zero the resolution parameter, which takes our solution to the KT one, or kill the number of fractional D3-branes. For this latter case, we showed that the effect of NATD was to change the $SU(3)$ structure background to an orthogonal $SU(2)$ structure.
We finally studied the central charge of the dual field theories prior to and after the NATD; we found that the central charges match up to a constant term, depending on the global details of the geometry, although a formal procedure to extract global topological information by which we can determine the bounds of the NATD coordinates is still outstanding.

\section*{Acknowledgments}
We would like to thank Carlos N\'u\~nez, Eoin \'O Colg\'ain, Niall Macpherson and Daniel Schofield for useful discussions during this project, and for suggestions for the direction of this paper. S. Zacar\'ias would like to thank the physics department of Swansea University, where this work was initiated, for their kind hospitality. The work of K.~S. Kooner is supported by an STFC scholarship and S. Zacar\'ias work was supported by a CONACyT-M\'exico scholarship and UG PIFI projects.

\section*{references}

\bibliography{bibliography}

\end{document}